\documentclass[twocolumn,showpacs,fleqn,nobibnotes]{revtex4}

\usepackage{amsmath}
\usepackage{graphicx}
\usepackage{float}
\usepackage{subfigure}

\def\lsim{\raise0.3ex\hbox{$<$\kern-0.75em\raise-1.1ex\hbox{$\sim$}}}
\def\gsim{\raise0.3ex\hbox{$>$\kern-0.75em\raise-1.1ex\hbox{$\sim$}}}

\newcommand{\rr}{\mbox{\boldmath $r$}}

\begin{document}

\title{Diffractive photoproduction of radially excited $\psi(2S)$ mesons in photon-Pomeron  reactions in PbPb collisions at the CERN LHC}
\author{M.B. Gay Ducati, M.T. Griep and M.V.T. Machado}

\affiliation{High Energy Physics Phenomenology Group, GFPAE  IF-UFRGS \\
Caixa Postal 15051, CEP 91501-970, Porto Alegre, RS, Brazil}

\begin{abstract}
In this work we investigate the photoproduction of radially excited vector mesons off nuclei in heavy ion relativistic collisions. In particular,  we analyze the exclusive photoproduction of $\psi(2S)$  off nuclei, evaluating the coherent and the incoherent contributions to that process. The theoretical framework used in the present analysis is the light-cone dipole formalism and predictions are done for PbPb collisions at the CERN-LHC energy of 2.76 TeV. The theoretical uncertainties are analyzed and comparison is also done to the recent  ALICE Collaboration data for the $\psi(1S)$ state photoproduction.

\end{abstract}

\maketitle

\section{Introduction}

An outstanding feature of diffractive photoproduction of mesons at the high energy regime is the possibility to investigate the Pomeron exchange. In such an energy domain hadrons and photons can be considered as color dipoles in the mixed light cone representation \cite{nik}. In particular, their transverse sizes are to be considered frozen during the interaction. Then, the scattering process is characterized by the color dipole cross section representing the interaction of those color dipoles with the target (protons or nuclei). The color evolution of the dipole cross section at small Bjorken-$x$ is given by the solution of a non-linear evolution equation. It is known for a long time that dipole sizes of magnitude $r\sim 1/\sqrt{m_V^2+Q^2}$ ($m_V$ is the vector meson mass) are probed by the $1S$ vector meson production amplitude \cite{nik}. As far as heavy quarks are concerned, the sufficiently large mass of quarkonium states makes the amplitude to be perturbatively calculable even at photoproduction region $Q^2\rightarrow 0$. The diffractive production of the $2S$ radially excited vector mesons, like $\psi(2S)$  and $\Upsilon(2S)$, is specially interesting due to the node effect \cite{Nemchik:1996pp}. It means a strong cancellation of dipole size contributions to the production amplitude from the region above and below the node position in the $2S$ radial wavefunction \cite{Nemchik:2000de}. This is the origin of the large suppression of the photoproduction of radially excited vector mesons $2S$ versus $1S$. It is an experimental fact that the ratio $\sigma (\psi^{\prime})/\sigma (\psi)\simeq  0.2 $ at DESY-HERA energies at $Q^2=0$ and the ratio is a $Q^2$-dependent quantity as the electroproduction cross sections are considered \cite{H1psi2}. The question has risen intense debate a long time ago \cite{Hoyer,Suzuki} and it was found for instance in Ref. \cite{Nemchik:1996cw} 
 that the combination of the  energy dependence of the dipole cross section and the node of the radial wavefunction of $2S$ states leads to an anomalous $Q^2$ and energy dependence of diffractive production of $2S$ vector mesons. In addition, such anomaly appears also in the $t$-dependence of the differential cross section of radially excited $2S$ light vector mesons \cite{Nemchik:2000dd}, which is in contradiction with the usual monotonical behavior of corresponding $1S$ states.

Here, we focus on the photoproduction of radially excited vector mesons off nuclei in heavy ion relativistic collisions. In particular,  we analyze the exclusive photoproduction of $\psi^{\prime}$  off nuclei, $\gamma A \rightarrow \psi(2S) X$, where for coherent scattering one has $X=A$ whereas for incoherent case $X=A^*$ with $A^*$ being an excited state of the $A$-nucleon system. The theoretical framework used in the present work is the light-cone dipole formalism. In such framework, the $c\bar{c}$ fluctuation of the incoming quasi-real photon interacts with the nucleus target via the dipole cross section and the result is projected on the wavefunction of the observed hadron. In the energy regime we are interested, the dipole cross section depends on the gluon distribution in the target and nuclear shadowing of the gluon distribution is expected to reduce it compared to a proton target. Moreover, theoretically at high energies  one
expects the transition of the regime described by the linear
dynamics, where only the parton emissions are considered, to a new
regime where the physical process of recombination of partons becomes
important in the parton cascade and the evolution is given by a
nonlinear evolution equation (for recent reviews on the topic see Ref. \cite{hdqcd}).  This regime is characterized by the
limitation on the maximum phase-space parton density that can be
reached in the hadron wavefunction, the so-called parton saturation. The
transition is specified  by a typical scale, the so called saturation scale $Q_{\mathrm{sat}}$  \cite{hdqcd}, which is energy dependent. 

Recently, the ALICE Collaboration has measured the diffractive $\psi(1S)$ vector meson production at a relatively large rapidity $y\simeq 3$ \cite{ALICE1}  and central rapidities \cite{ALICE2} as well in the $\sqrt{s}=2.76$ TeV run, which opens the possibility of investigating small-$x$ physics with heavy nuclei. In addition, the incoherent $\psi (1S)$ cross section has been also measured \cite{ALICE2}.  This is interesting, as the saturation is enhanced for nuclear targets, i.e.  $Q_{\mathrm{sat}}\propto A^{1/3}$. The LHCb Collaboration has also measured the cross section in proton-proton collisions at $\sqrt{s}=7$ TeV  of exclusive dimuon final states, including the $\psi (2S)$ state \cite{LHCb}. The ratio at forward rapidity $2.0\leq \eta_{\mu^{\pm}}\leq 4.5$ in that case is $\sigma(\psi(2S))/\sigma(\psi(1S))= 0.19\pm 0.04$, which is still consistent to the color dipole approach formalism. Therefore, an investigation on the $\psi(2S)$ photoproduction in PbPb collisions at the LHC is interesting by itself and timely.

This paper is organized as follows. In next section we present a brief review of the diffractive photoproduction of vector mesons in electromagnetic reactions in nucleus-nucleus collisions focusing on the PbPb reactions at the LHC energy regime. In Section \ref{resultados} we  show our predictions for the $\psi(2S)$ photoproduction cross section including the coherent and incoherent contributions.  We also check the compatibility with the recent measurements of the $\psi (1S)$ state \cite{ALICE1,ALICE2}.  Moreover, we compare the current results to related approaches available in the literature. Finally, in Section \ref{conc} we summarize our main results and conclusions.

\section{Photon-pomeron process in relativistic nucleus-nucleus collisions}
\label{coerente}

The electromagnetic interaction is dominant in the nucleus-nucleus interaction at large impact parameter and at ultra relativistic energies. In  heavy ion colliders, the heavy nuclei give rise to strong electromagnetic fields due to the coherent action of all protons in the nucleus, which can interact with each other. Accordingly, the total cross section for a given process can be factorized in terms of the equivalent flux of photons of the hadron projectile and  the photon-photon or photon-target production cross section \cite{upcs}.  In what follows our main focus shall be in photon - hadron processes which is relevant for the photoproduction of radially excited vector mesons.
Considering the requirement that  photoproduction
is not accompanied by hadronic interaction an analytic expression for the equivalent photon flux of a nuclei can be calculated \cite{upcs}
\begin{eqnarray}
\frac{dN_{\gamma}\,(\omega)}{d\omega} & = & \frac{2\,Z^2\alpha_{em}}{\pi\,\omega}\, \left[\xi_R^{AA}\,K_0\,(\xi_R^{AA})\, K_1\,(\xi_R^{AA}) \right. \nonumber \\
& - & \left. \frac{(\xi_R^{AA})^2}{2}\,K_1^2\,(\xi_R^{AA})-  K_0^2\,(\xi_R^{AA}) \right].
\label{fluxint}
\end{eqnarray}
where  $\omega$ is the photon energy,  $\gamma_L$ is the Lorentz boost  of a single beam and $K_0(\xi)$ and  $K_1(\xi)$ are the
modified Bessel functions.
Considering symmetric nuclei having radius $R_A$, one has $ \xi_R^{AA}=2R_A\omega/\gamma_L$. 

The cross section for the photoproduction of $\psi^{\prime}$ off nuclei in heavy ion relativistic collisions is  given by,
\begin{eqnarray}
\sigma (AA \rightarrow \psi(2S) X) = \int \limits_{\omega_{min}}^{\infty} d\omega \int dt \,\frac{dN_{\gamma}(\omega)}{d\omega}\,\frac{d\sigma}{dt} \left(W_{\gamma A},t\right),
\label{sigAA}
\end{eqnarray}
where $\frac{d\sigma}{dt}$ is the differential cross section for the process    $\gamma A \rightarrow \psi^{\prime} X$, $\omega_{min}=M_{\psi}^2/4\gamma_L m_p$, $W_{\gamma p}^2=2\,\omega\sqrt{S_{\mathrm{NN}}}$  and
$\sqrt{S_{\mathrm{NN}}}$ is  the c.m.s energy of the
nucleus-nucleus system. Since photon emission is coherent over the entire nucleus and the photon is colorless we expect that the events to be characterized by  $X = A$ and two   rapidity gaps in the case of coherent process. In the incoherent process, $X=A^*$ (excited nucleus state) as already mentioned.

The rapidity distribution $y$ for quarkonium photoproduction in nucleus-nucleus collisions can be also computed  directly from Eq. (\ref{sigAA}), by using its  relation with the photon energy $\omega$, i.e. $y\propto \ln \, (2 \omega/m_X)$.  Explicitly, the rapidity distribution is written down as,
\begin{eqnarray}
\frac{d\sigma \,\left[A A \rightarrow   A\otimes \psi(2S) \otimes X \right]}{dy} = \omega \frac{dN_{\gamma} (\omega )}{d\omega }\,\sigma_{\gamma A \rightarrow \psi(2S) X }\left(\omega \right),
\label{dsigdy}
\end{eqnarray}
where $\otimes$ represents the presence of a rapidity gap.
Consequently, given the photon flux, the rapidity distribution is thus a direct measure of the photoproduction cross section for a given energy.

Let us consider photon-nucleus scattering in the light-cone dipole frame, in which most of the energy is
carried by the hadron, while the  photon  has
just enough energy to dissociate into a quark-antiquark pair
before the scattering. In this representation the probing
projectile fluctuates into a
quark-antiquark pair (a dipole) with transverse separation
$\rr$ long after the interaction, which then
scatters off the hadron \cite{nik}.
In the dipole picture the   amplitude for vector meson  production off nucleons reads  as (See e.g. Refs. \cite{nik,mesons})
\begin{eqnarray}
\, {\cal A}\, (x,Q^2,\Delta)  = \sum_{h, \bar{h}}
\int dz\, d^2\rr \,\Psi^\gamma_{h, \bar{h}}\,{\cal{A}}_{q\bar{q}} \, \Psi^{V*}_{h, \bar{h}} \, ,
\label{sigmatot}
\end{eqnarray}
where $\Psi^{\gamma}_{h, \bar{h}}(z,\,\rr,Q^2)$ and $\Psi^{V}_{h,
  \bar{h}}(z,\,\rr)$ are the light-cone wavefunctions  of the photon  and of the  vector meson, respectively. The
   quark and antiquark helicities are labeled by $h$ and $\bar{h}$, variable $\rr$ defines the relative transverse
separation of the pair (dipole), $z$ $(1-z)$ is the
longitudinal momentum fractions of the quark (antiquark). The quantity  $\Delta$ denotes the transverse momentum lost by the outgoing proton ($t = - \Delta^2$) and $x$ is the Bjorken variable. Moreover, ${\cal{A}}_{q\bar{q}}$ is the elementary amplitude for the scattering of a dipole of size $\rr$ on the target. In a compact notation, the non-forward amplitude and the differential cross section for exclusive production of charmonia (or other final state)  off a nucleon target, respectively, are given by,
\begin{eqnarray}
 {\cal A}\, (x,Q^2,\Delta) & = & \langle \Psi^V|{\cal{A}}_{q\bar{q}}(x,\rr,\Delta)|\Psi^{\gamma}\rangle
\label{coherent} \\
\frac{d\sigma(s,Q^2)}{dt} & = & \frac{1}{16\pi } |{\cal A}\, (x,Q^2,\Delta)|^2
\label{incoherent}
\end{eqnarray} 

In the numerical calculation shown in next section, the corrections due to skewedness effect (off-diagonal gluon exchange) and real part of amplitude are also taken into account. Detail on the model dependence on these corrections can be found for instance in Ref. \cite{GM}.

The photon wavefunctions appearing in Eq. (\ref{sigmatot}) are relatively well known \cite{mesons}. concerning the meson wavefunction, in the current calculation we consider the boosted gaussian wavefunction:
\begin{eqnarray}
\psi_{\lambda, h\bar{h}}^{nS}  & = & \sqrt{\frac{N_c}{4\pi}} \frac{\sqrt{2}}{z(1-z)} \Biggl \{ \delta_{h,\bar{h}} \delta_{\lambda,2h} m_c +  i(2h)\delta_{h,-\bar{h}} e^{i\lambda\phi} \nonumber \\
&\times &\left[ (1-z)\delta_{\lambda,-2h} + z\delta_{\lambda,2h}   \right]\partial_r \Biggr \} \,\phi_{nS}(z,r). 
\end{eqnarray}

Here, $\phi(z,r)$ in the mixed $(r,z)$ representation is obtained by boosting a Schr\"{o}dinger gaussian wavefunction in momentum representation, $\Psi(z,{\bf{k}})$. In this case, one obtains the following expression for the $1S$ state \cite{Sandapenpsi}: 
\begin{eqnarray}
\phi_{1S}(r,z) &=& N_T^{(1S)}\Biggl \{ 4z(1-z) \sqrt {2\pi R_{1S}^{2}}\exp\left[-{m_{q}^{2}R_{1S}^{2} \over 8z(1-z)}\right] \nonumber \\
& \times & \exp\left[-{2z(1-z)r^{2} \over R_{1S}^{2}}\right]
\exp\left[{m_{q}^{2}R_{1S}^{2} \over 2}\right] \Biggr \},
\label{eq:wf1S}
\end{eqnarray}
where for the $1S$ ground state vector mesons we determine the parameters
$R_{1S}^{2}$ and $N_T$ by considering the normalization property of wavefunctions and the predicted decay widths.

The radial wavefunction of the $\psi(2S)$ is obtained by the following modification of the $1S$ state \cite{Nemchik:1996pp}:
\begin{eqnarray}
\phi_{2S}(r,z) &=& N_T^{(2S)}\Biggl \{ 4z(1-z) \sqrt {2\pi R_{2S}^{2}}
\exp\left[-{m_{q}^{2}R_{2S}^{2} \over 8z(1-z)}\right] \nonumber \\
& \times & \exp\left[-{2z(1-z)r^{2} \over R_{2S}^{2}}\right]  
\exp\left[{m_{q}^{2}R_{2S}^{2} \over 2}\right] \left(1-\hat{\phi} \right)  \Biggr \} ,\nonumber \\
 \hat{\phi} &= & \alpha\,\left[1 + m_{q}^{2}R_{2S}^{2} -
{m_{q}^{2}R_{2S}^{2} \over 4z(1-z)}
+ {4z(1-z) \over R_{2S}^{2}}r^{2} \right] 
, \nonumber \\
\label{eq:wf2S}
\end{eqnarray}
 with the new parameter $\alpha$ controlling the position of the node. In addition, the two parameters $\alpha$ and $R_{2S}$ are determined from the
orthogonality conditions for the meson wavefunction. See, for instance, the example for the determination of parameters for the Upsilon photoproduction in Ref.  \cite{SandaUps}. 

At this point some comments are in order. First, we are using a particular choice for the meson wavefunctions, Eqs. \ref{eq:wf1S} and \ref{eq:wf2S}. The boosted gaussian wavefunction considered here is a simplification of the NNPZ wavefunction presented in Refs. \cite{nik,Nemchik:1996pp}. It has been compared to recent analysis of DESY-HERA data for vector meson exclusive processes. For instance, in Ref. \cite{Sandapenpsi} the boosted gaussian wavefunction was successfully compared to the light mesons and $J/\psi$ production data. In Ref. \cite{SandaUps}, the production of $\Upsilon (1S)$ and its excited states ($\Upsilon (2S)$, $\Upsilon (3S)$) was investigated and very good agreement with  DESY-HERA data was found. Another point to be analyzed is the role played by the node effect to describe the measured ratio $\sigma (\psi^{\prime})/\sigma (J/\psi)$ in the photoproduction case. Such a ratio is sensitive to the time-scale of the production process. In the dipole approach the interactions occur during the period where the color dipole is compact having a transverse size $r \simeq 1/m_q$ and the production cross section is proportional to the square of the quarkonium wavefunction at origin, $\sigma \propto |\phi (0)|^2$. On the other hand, further interactions depend on the wavefunction profile for transverse sizes larger than $r_B = {\cal O}(1/\alpha_s m_q)$, the so-called Bohr radius. In exclusive charmonia electroproduction at relatively large $Q^2$ the dipole size is of order $1/Q^2\ll r_B$ and the cross section is predicted to be proportional to $|\phi_n (0)|^2$. This leads the ratio to be of order $|\phi_{2S} (0)|^2/|\phi_{1S} (0)|^2\simeq 0.6$ at large $Q^2$ whereas the measured value in photoproduction is around 0.16 \cite{H1psi2}.  It has been determined long time ago in Refs. \cite{nik,Nemchik:1996pp} that the moderate value of charm mass and the dominant color transparency behavior of dipole cross section $\sigma_{dip}\propto r^2$ imply the amplitude to probe the meson wavefunction at a transverse size around $r_B$. This fact reduces the $\psi (2S)$ contribution due to the node in its wavefunction and correctly describes the measured DESY-HERA ratio. Along these lines, it was explicitly shown in Ref. \cite{Suzuki} the at $Q^2\rightarrow 0$ the leading logarithmic approximation $rl_{\perp} \ll 1$, which gives the usual $\sigma_{dip}\propto r^2$, is not able to provide alone the correct value for the ratio $\psi^{\prime}/\psi$. Here, $l_{\perp}$ is the exchanged gluon transverse momentum in a two-gluon exchange model. Therefore, important contributions come from the overlap of the large-sized color dipole  configurations and the $\psi (2S)$ wavefunction. Thus, despite the leading logarithmic approximation to be able to describe the $J/\psi$ production cross section the same is not true for the excited states as the  $\psi^{\prime}$. This is the reason why we will use a model for the dipole cross section that takes into account the correct  behavior for  large dipole configurations (the transition hard-soft is given by the saturation scale). 

The exclusive $\psi(2S)$ photoproduction off nuclei for coherent and incoherent processes can be simply computed in high energies where the large coherence length $l_c\gg R_A$ is fairly valid. In such case the transverse size of $c\bar{c}$ dipole is frozen by Lorentz effects. The expressions for the coherent and incoherent cross sections are given by \cite{Boris},
\begin{eqnarray}
\sigma_{coh}^{\gamma A} & = & \int d^2b\, |\langle \Psi^V|1-\exp\left[-\frac{1}{2}\sigma_{dip}(x,\rr) T_A(b)  \right]|\Psi^{\gamma}\rangle |^2, \label{eq:coher} \nonumber \\
 \\
\sigma_{inc}^{\gamma A} & = & \frac{1}{16\pi\,B_V(s)}\int d^2b\,T_A(b) \nonumber \\
 & \times & |\langle \Psi^V|\sigma_{dip}(x,\rr) \exp\left[-\frac{1}{2}\sigma_{dip}(x,\rr)T_A(b)  \right]|\Psi^{\gamma}\rangle|^2. \nonumber \\
\label{eq:incoh}
\end{eqnarray} 
where $T_A(b)= \int dz\rho_A(b,z)$  is the nuclear thickness function given by integration of nuclear density along the trajectory at a given impact parameter $b$. In addition, $B_V$ is the diffractive slope parameter in the reaction $\gamma^*p\rightarrow \psi p$. Here, we consider the energy dependence of the slope using the Regge motivated expression $B_V(W_{\gamma p})=b_{el}^V + 2\alpha^{\prime}\log \frac{W_{\gamma p}^2}{W_0^2}$ with $\alpha^{\prime}=0.25$ GeV$^{-2}$ and $W_0=95$ GeV.  It is used the measured slopes \cite{H1psi2} for $\psi(1S)$ and $\psi(2S)$ at $W_{\gamma p}=90$ GeV, i.e. $b_{el}^{\psi(1S)}= 4.99\pm 0.41$ GeV$^{-2}$ and $b_{el}^{\psi(2S)}= 4.31\pm 0.73$ GeV$^{-2}$, respectively.

The last ingredient is the model for the dipole cross section in Eqs. (\ref{eq:coher}) and (\ref{eq:incoh}). In our calculation, we consider the Color Glass Condensate model \cite{IIM} for $\sigma_{dip}(x,r)$. This model has been tested for a long period against DIS, diffractive DIS and exclusive production processes in $ep$ collisions. In addition, we allow for its  renormalization by the effect of gluon shadowing phenomenon as the gluon density in nuclei at small-$x$ region is known to be suppressed compared to a free nucleon. That is, we will take $\sigma_{dip}\rightarrow R_G(x,Q^2,b)\sigma_{dip}$  following studies in Ref. \cite{Borispsi}. The factor $R_G$ is the nuclear gluon density ratio. In the present investigation we will use the nuclear ratio from the leading twist theory of nuclear shadowing based on generalization of the Gribov-Glauber multiple scattering formalism as investigated in Ref. \cite{GFM}. We used the two models available for $R_G(x,Q^2)$ in \cite{GFM}, Models 1 and 2, which correspond to higher nuclear shadowing and lower nuclear shadowing, respectively. Such a choice is completely arbitrary and other nuclear  gluon ratios available in literature could be considered. It would be also interesting to investigate the effect of using the impact parameter dependent nuclear parton distribution ratios. We discuss about this distinct issues in next section.

\begin{figure}[t]
\includegraphics[scale=0.45]{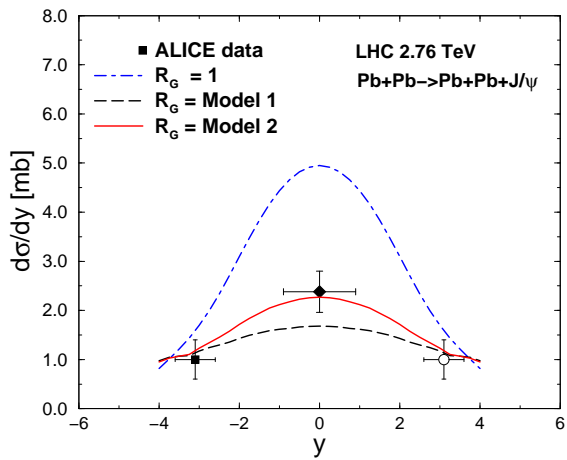}
\caption{(Color online) \it The rapidity distribution of coherent $\psi(1S)$ meson photoproduction at $\sqrt{s}=2.76$ TeV in PbPb collisions at the LHC. The theoretical curves stand for color dipole formalism using $R_G=1$ (dot-dashed curve) and two scenarios for the nuclear gluon distribution (solid and long-dashed curves, see text). Data from ALICE collaboration \cite{ALICE1,ALICE2}.}
\label{fig:1}
\end{figure}

\section{Results and discussions}
\label{resultados}

Let us start by checking the present theoretical approach against the recent data for the $1S$ state measured by ALICE Collaboration at the energy of 2.76 TeV in PbPb collisions at the LHC \cite{ALICE1,ALICE2}. In Fig. \ref{fig:1} we present the numerical calculations for the rapidity distribution of coherent $\psi(1S)$ state within the color dipole formalism, Eqs. (\ref{dsigdy}) and (\ref{eq:coher}), using distinct scenarios for the nuclear gluon shadowing. The dot-dashed curve represents the result using $R_G=1$ and it is consistent with previous calculations using the same formalism \cite{GM}. It overestimates the ALICE data on the backward (forward) and mainly in central rapidities. In the backward/forward rapidity  case, the overestimation is already expected as a proper threshold factor for $x\rightarrow 1$ was not included in the present calculation. In that kinematical region either a small-$x$ photon scatters off a large-$x$ gluon or vice-versa. For instance, for $y\simeq \pm 3$ one gets $x$ large as $0.02$. On the other hand, for central rapidity $y=0$ one can be obtained $x=M_Ve^{\pm y}/\sqrt{s_{\mathrm{NN}}}$ smaller than $10^{-3}$ for the nuclear gluon distribution. In such a case, considering $R_G=1$ the ALICE data \cite{ALICE2} is overestimate by a factor 2 or so, as already noticed in recent study of Ref. \cite{LM}. The situation is improved if we consider nuclear shadowing renormalising the dipole cross section. The reason is that the gluon density in nuclei at small Bjorken $x$ is expected to be suppressed compared to a free nucleon due to interferences. For the ratio of the gluon density, $R_G(x,Q^2=m_V^2/4)$, we have considered the theoretical evaluation of Ref. \cite{GFM}. There, two scenarios for the gluon shadowing are investigated: Model 1 corresponds to a strong gluon shadowing and Model 2 concerns to small nuclear shadowing. The consequence of renormalizing the dipole cross section by gluon shadowing effects is represented  by the long-dashed (Model 1) and solid (Model 2) lines, respectively. Clearly, the small shadowing option is preferred in the current analysis. It is worth to mention that the theoretical uncertainty related to the choice of meson wavefunction is relatively large. As a prediction at central rapidity, one obtains $\frac{d\sigma}{dy} (y=0) = 4.95, \,1.68$ and $2.27$ mb for calculation using $R_G=1$, Model 1 and Model 2, respectively. Here, a word of caution is needed as we are considering $R_G$ as independent on the impact parameter. It is long time known that a $b$-dependent ratio could give a smaller suppression compared to presented in our calculation. For instance, in Ref. \cite{Borispsi} the suppression is of order 0.85 for the LHC energy and central rapidity. 

\begin{figure}[t]
\includegraphics[scale=0.45]{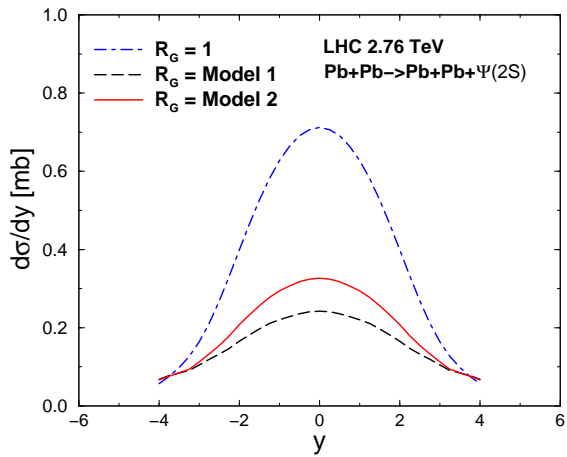}
\caption{(Color online) \it The rapidity distribution of coherent $\psi(2S)$ meson photoproduction at $\sqrt{s}=2.76$ TeV in PbPb collisions at the LHC. The theoretical curves follow the same notation as in the previous figure.}
\label{fig:2}
\end{figure}

In Fig. \ref{fig:2} we show our predictions for the coherent photoproduction of $\psi(2S)$ state. This is the first estimate in literature for the photoproduction of $2S$ state in nucleus-nucleus collisions. The theoretical predictions follow the general trend as for the $1S$ state, where the notation for the curves are the same as used in Fig. \ref{fig:1}. In particular, for $R_G=1$ one obtains for central rapidity $\frac{d\sigma}{dy}(y=0) = 0.71$ mb and the following in the forward/backward region  $\frac{d\sigma}{dy}(y=\pm 3) = 0.16$ mb. When introducing the suppression in dipole cross section due nuclear shadowing one gets instead $\frac{d\sigma}{dy}(y=0) = 0.24$ mb and  0.33 mb  for Model 1 and Model 2, respectively. At central rapidities, the meson state ratio is evaluated to be $R_{\psi}^{y=0}=\frac{\sigma_{\psi(2S)}}{dy}/\frac{d\sigma_{\psi(1S)}}{dy} (y=0)= 0.14$ in case $R_G=1$ which is consistent with the ratio measured in CDF, i.e. $0.14\pm 0.05$, on the observation of exclusive charmonium production at 1.96 TeV in $p\bar{p}$ collisions \cite{CDF}. A similar ratio is obtained using Model 1 and Model 2 at central rapidity as well. As a prediction for the planned LHC run in PbPb mode at 5.5 TeV, we obtain $\frac{d\sigma_{coh}}{dy}(y=0) = 1.27$ mb and $\frac{d\sigma_{inc}}{dy}(y=0) = 0.27$ mb  for the coherent and incoherent $\psi(2S)$ cross sections (upper bound using $R_G=1$), respectively.

\begin{figure}[t]
\includegraphics[scale=0.45]{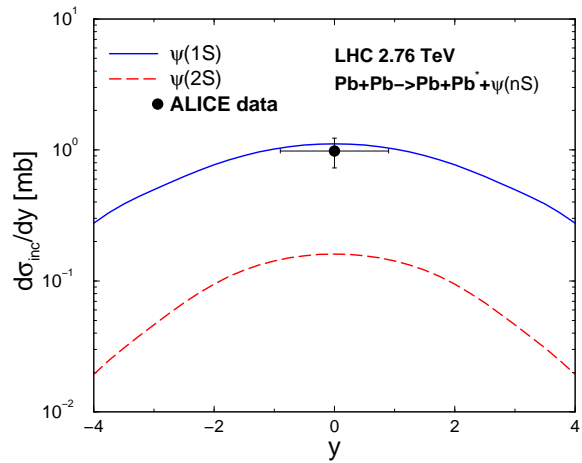}
\caption{(Color online) \it The rapidity distribution of incoherent $\psi(1S)$ (solid line) and $\psi(2S)$ (dashed line) meson photoproduction at $\sqrt{s}=2.76$ TeV in PbPb collisions at the LHC. Data from ALICE collaboration \cite{ALICE2}.}
\label{fig:3}
\end{figure}

Finally, in Fig. \ref{fig:3} we show the incoherent contribution to the rapidity distribution for both $\psi(1S)$ (solid line) and $\psi (2S)$ (dashed line) meson states. The theoretical estimates are done using Eq. (\ref{incoherent}) taking into account the corresponding diffractive slope for each meson state as discussed in the previous section.  For the $\psi(1S)$ state, the present calculation can be directly compared with those studies presented in Ref. \cite{LM}. It was found in \cite{LM} that the incoherent cross section  $\frac{d\sigma_{\mathrm{inc}}}{dy}$  ranges between 0.5 to 0.7 mb (using IIM dipole cross section) or between 0.7 to 0.9 mb (using fIPsat dipole cross section) at central rapidities, with the uncertainty determined by the distinct meson wavefunction considered. In our case, we obtained  $\frac{d\sigma_{\mathrm{inc}}}{dy}(y=0) = 1.1$ mb using a different expression for the incoherent amplitude, Eq. (\ref{eq:incoh}). Our result fairly describes the recent ALICE data \cite{ALICE2} for the incoherent cross section at mid-rapidity, $\frac{d \sigma_{inc}^{\mathrm{ALICE}}}{dy} (-0.9<y<0.9) = 0.98 \pm 0.25$ mb. As a prediction for the $\psi(2S)$ state, we have found $\frac{d\sigma_{\mathrm{inc}}}{dy} = 0.16$ mb for central rapidities. In both cases we have only computed the case for $R_G=1$. Therefore, this gives an upper bound for the incoherent cross section compared to Model 1 and Model 2 calculation. We notice that for the incoherent case, the gluon shadowing is weaker than the coherent case and the reduction is around 20 \% compared to the case $R_G=1$. As expected, the incoherent  piece is quite smaller compared to the main coherent contribution. As an example of order of magnitude, the ratio incoherent/coherent is a factor 0.22 for the $1S$ state and 0.23 for the $2S$ state at central rapidity. 

\section{Summary}
\label{conc}

We have investigated the photoproduction of radially excited vector mesons off nuclei in heavy ion relativistic collisions as the $\psi(2S)$ charmonium state. The theoretical framework used in the present analysis is the light-cone dipole formalism and predictions are done for PbPb collisions at the CERN-LHC energy of 2.76 TeV. The effect of suppressing of the dipole cross section due to the gluon shadowing was studied and the results for $R_G=1$ give the larger cross sections. It was found that the coherent exclusive photoproduction of $\psi(2S)$  off nuclei has an upper bound of order 0.71 mb at $y=0$ down to 0.10 mb for backward/forward rapidities $y=\pm 3$. The incoherent contribution was also computed and it is a factor 0.2 below the coherent one. Comparison has been done to the recent  ALICE Collaboration data for the $\psi(1S)$ state photoproduction and the analysis shows that a small nuclear shadowing $R_G(x,Q^2=\frac{m_V^2}{4})$ is preferred in data description whereas the usual $R_G=1$ value overestimates the central rapidity cross section by a factor 2. On the other hand, the present theoretical approach fairly describes the ALICE data for incoherent cross section. Thus, the central rapidity data measured by ALICE Collaboration for the rapidity distribution of the $\psi(1S)$ state is crucial to constrain the nuclear gluon function. The cross section for exclusive quarkonium  production is proportional to $[\alpha(Q^2)xg_A(x,Q^2)]^2$ in the leading-order pQCD calculations, evaluated at the relevant scale $Q^2\approx m_V^2/4$ and at momentum fraction $x\simeq 10^{-3}$ in central rapidities. The theoretical uncertainty is large and it has been investigated in several studies \cite{AGG,AB}. Along these line, the authors of Ref. \cite{GKSZ} extract the nuclear suppression factor, $S(x\approx 10^{-3})=0.61 \pm 0.064$, using the ALICE data on coherent $\psi (1S)$ and considering the nuclear gluon shadowing predicted by nuclear pdf's and by leading twist nuclear shadowing.

\begin{acknowledgments}
This work was  partially financed by the Brazilian funding
agencies CNPq and FAPERGS and by the French-Brazilian scientific cooperation project CAPES-COFECUB 744/12. MVTM thanks to Magdalena Malek, Heikki M\"{a}ntysaari and Daniel Tapia Takaki for helpful comments.
\end{acknowledgments}


\begin{thebibliography}{99}

\bibitem{nik} N. N. Nikolaev, B. G. Zakharov,  Phys. Lett. B  {\bf 332}, 184 (1994); {Z. Phys. C} {\bf 64}, 631 (1994).

\bibitem{Nemchik:1996pp} 
  J.~Nemchik, N.~N.~Nikolaev, E.~Predazzi and B.~G.~Zakharov,
  Phys.\ Lett.\ B {\bf 374}, 199 (1996).

\bibitem{Nemchik:2000de} 
  J.~Nemchik,
  Phys.\ Rev.\ D {\bf 63}, 074007 (2001)

\bibitem{H1psi2} C. Adloff {\it et al.} [H1 Collaboration], Phys. Lett. B541, 251 (2002).

\bibitem{Hoyer} P.~Hoyer and S.~Peigne,  Phys.\ Rev.\ D {\bf 61}, 031501 (2000).

\bibitem{Suzuki} K.~Suzuki, A.~Hayashigaki, K.~Itakura, J.~Alam and T.~Hatsuda,  Phys.\ Rev.\ D {\bf 62}, 031501 (2000).

\bibitem{Nemchik:1996cw} 
  J.~Nemchik, N.~N.~Nikolaev, E.~Predazzi and B.~G.~Zakharov,
  Z.\ Phys.\ C {\bf 75}, 71 (1997)

\bibitem{Nemchik:2000dd} 
  J.~Nemchik,
  Eur.\ Phys.\ J.\ C {\bf 18}, 711 (2001)

\bibitem{hdqcd} 
  F.~Gelis, E.~Iancu, J.~Jalilian-Marian and R.~Venugopalan,
    Ann.\ Rev.\ Nucl.\ Part.\ Sci.\  {\bf 60}, 463 (2010);
  H.~Weigert,  Prog.\ Part.\ Nucl.\ Phys.\  {\bf 55}, 461 (2005); J.~Jalilian-Marian and Y.~V.~Kovchegov, Prog.\ Part.\ Nucl.\ Phys.\  {\bf 56}, 104 (2006).

\bibitem{ALICE1}  B. Abelev {\it et al.} [ALICE Collaboration], Phys. Lett. B718, 1273 (2013).

\bibitem{ALICE2}  E. Abbas {\it et al.} [ALICE Collaboration], arXiv:1305.1467 [nucl-ex].


\bibitem{LHCb} R. Aaij {\it et al.}  [LHCb Collaboration],
  J.\ Phys.\ G {\bf 40}, 045001 (2013).

\bibitem{upcs}
 G. Baur, K. Hencken, D. Trautmann, S. Sadovsky, Y. Kharlov, Phys.
Rep. {\bf 364}, 359 (2002);
 C.~A. Bertulani, S.~R.~Klein and J.~Nystrand, Ann. Rev. Nucl. Part. Sci. {\bf 55}, 271 (2005).

\bibitem{mesons}
 A.~C.~Caldwell and M.~S.~Soares,
  Nucl.\ Phys.\  A {\bf 696}, 125 (2001); H.~Kowalski and D.~Teaney,
  Phys.\ Rev.\  D {\bf 68}, 114005 (2003);  J. R. Forshaw, R. Sandapen and G. Shaw, Phys. Rev. D69
, 094013 (2004);  C.~Marquet, R.~Peschanski and G.~Soyez,
  Phys.\ Rev.\  D {\bf 76}, 034011 (2007);
H.~Kowalski, L.~Motyka and G.~Watt,
  Phys.\ Rev.\  D {\bf 74}, 074016 (2006).

\bibitem{GM} 
  V.~P.~Goncalves and M.~V.~T.~Machado,
  Phys.\ Rev.\ C {\bf 84}, 011902 (2011).

\bibitem{Sandapenpsi}
J.~R.~Forshaw, R.~Sandapen and G.~Shaw,
  JHEP {\bf 0611}, 025 (2006).

\bibitem{SandaUps} B.E. Cox, J.~R.~Forshaw and  R.~Sandapen,
  JHEP {\bf 0906 }, 034 (2009).


\bibitem{Boris} B. Z. Kopeliovich and B. G. Zakharov, Phys. Rev. D 44, 3466 (1991). 

 
\bibitem{IIM}
  E.~Iancu, K.~Itakura and S.~Munier,
  Phys.\ Lett.\ B {\bf 590}, 199 (2004).

\bibitem{Borispsi} Y.~.P.~Ivanov, B.~Z.~Kopeliovich, A.~V.~Tarasov and J.~Hufner,
  Phys.\ Rev.\ C {\bf 66}, 024903 (2002).

\bibitem{GFM} 
  L.~Frankfurt, V.~Guzey and M.~Strikman,
  Phys.\ Rept.\  {\bf 512}, 255 (2012)


\bibitem{LM} 
  T.~Lappi and H.~Mantysaari,
Phys. Rev. C {\bf 87}, 032201 (2013).  

\bibitem{CDF} T. Aaltonen {\it et al.} [CDF Collaboration],  Phys. Rev. Lett. {\bf 102}, 242001 (2009).


\bibitem{AGG} A.L. Ayala Filho, V.P. Gon\c{c}alves and M.T. Griep, Phys. Rev. C {\bf 78}, 044904 (2008).

\bibitem{AB} A. Adeluyi and C.A. Bertulani, Phys. Rev. C {\bf 85}, 044904 (2012).

\bibitem{GKSZ} V. Guzey, E. Kryshen, M. Strikman and M. Zhalov, arXiv:1305.1724 [hep-ph].

\end{thebibliography}
\end{document}